\documentclass[nonacm, sigchi]{acmart}

\AtBeginDocument{%
  \providecommand\BibTeX{{%
    \normalfont B\kern-0.5em{\scshape i\kern-0.25em b}\kern-0.8em\TeX}}}
\usepackage[ruled,vlined, linesnumbered]{algorithm2e}

\begin{document}

%
\title{Beyond Next Item Recommendation: Recommending and Evaluating List of Sequences}

%
\author{Makbule Gulcin Ozsoy}
\email{makbule.ozsoy@ucd.ie}
\affiliation{%
}






%
\renewcommand{\shortauthors}{MGOzsoy}

%
\begin{abstract}
Recommender systems (RS) suggest items-based on the estimated preferences of users. Recent RS methods utilise vector space embeddings and deep learning methods to make efficient recommendations. However, most of these methods overlook the sequentiality feature and consider each interaction, e.g., check-in, independent from each other. The proposed method considers the sequentiality of the interactions of users with items and uses them to make recommendations of a list of multi-item sequences. 
The proposed method uses FastText~\cite{bojanowski2016enriching}, a well-known technique in natural language processing (NLP), to model the relationship among the subunits of sequences, e.g., tracks, playlists, and utilises the trained representation as an input to a traditional recommendation method. The recommended lists of multi-item sequences are evaluated by the ROUGE~\cite{lin2003automatic,lin2004rouge} metric, which is also commonly used in the NLP literature.
The current experimental results reveal that it is possible to recommend a list of multi-item sequences, in addition to the traditional next item recommendation. Also, the usage of FastText, which utilise sub-units of the input sequences, helps to overcome cold-start user problem.
Even though current experimental results are promising, there are many missing pieces in the experimental section. In the future, I want to analyse and execute additional experiments.

\end{abstract}

%
%


%

%
\maketitle

\section{Introduction} \label{intro}


Recommender systems~(RS) suggest items-based on the estimated preferences of users \cite{MassaA07, TavakolifardA12}. Recent literature in RS frequently use vector space embeddings and deep learning methods to make efficient recommendations~\cite{grbovic2015commerce, musto2018deep, ai2018learning,yang2018unsupervised}. However, most of these works do not use the sequentiality of interactions and consider each interaction, e.g., check-in, independent from each other~\cite{zhao2016gt, guo2018exploiting}.

In natural language processing (NLP), sequentiality among subunits naturally occurs. The word ordering in a sentence can reveal various information or affect the emphasis of the different part of the sentences. For example, when we read an adjective, such as' black', we naturally expect to see a noun,e.g., 'bird', rather than a verb, e.g., 'run'. 
Many vector space embedding models, such as Word2Vec \cite{MikolovCoRR13, MikolovSCCD13}, FastText~\cite{bojanowski2016enriching} or recent transformer models, use the sequential information among the textual units, e.g. words, n-grams. 
Inspiring from NLP applications, this work adapts FastText method to the RS domain by making an analogy in between textual data and user-item interactions data.

The proposed method uses the sequentiality information on the interaction of users with items as the input. For example, a user's check-ins during a trip form a sequence, and this sequence is used as an input. Then, the input sequences are grouped into subsequences (or the session information is used, if available) and used for training the FastText model. This model reveals the relationship among items. Finally, these models are used in a more traditional setting, e.g. content filtering, to decide on the sequences to recommend. Unlike more traditional next-item recommendation, the proposed method is capable of recommending a list of multi-item sequences. For example, it can recommend music playlists, each of which is composed of individual music tracks.

Another aspect that needs further research is the evaluation of the  recommended sequences. In the RS literature, there are several evaluation metrics measuring the performance of recommendation algorithms on several different aspects, such as RMSE, classification accuracy, NDCG, coverage, novelty, diversity~\cite{gunawardana2015evaluating}. These metrics are already extended and adapted for evaluating a sequence-based recommender system~\cite{monti2019sequeval}. However, these methods, even the extended versions, consider the sequences as a list of items. They discard the order among the recommended items~\cite{monti2019sequeval}. This evaluation problem is also observed in NLP. In NLP, there are metrics for comparing overlapping textual units between the automatically generated outputs and the ideal results created by humans~\cite{lin2003automatic,lin2004rouge}. In this work, we adopt Recall Oriented Understudy for Gisting Evaluation (ROUGE) metric~\cite{lin2003automatic,lin2004rouge}, which is originally used in text summarization domain, to the RS domain for the evaluation of the multi-item sequence recommendations. 

In the upcoming sections, the FastText method and the ROUGE metric are detailed, and how they are adapted to the RS domain for making sequence recommendations are explained. Then the experimental setting and the evaluation results are presented. Finally, the paper is concluded.

\section{Related Work}\label{relWork}
In the literature, various approaches are employed to make recommendations, from traditional methods like collaborative filtering, content-based filtering, matrix factorization to more recent vector space embeddings, deep learning methods \cite{pan2008one, Ye2010, Ma:2011:RSS:1935826.1935877, GeorgievN13, Ozsoy14, li2015rank, ZhangWang:2015, HeLLSC16, musto2018deep, ai2018learning}. 

Collaborative filtering and content-based filtering methods use item-user or user-user similarities. Example collaborative filtering-based recommendation methods belong to Ye et al.~\cite{Ye2010}, Yuan et al.~\cite{Yuan:2013:TPR:2484028.2484030}, Zhang and Wang\cite{ZhangWang:2015} and Ozsoy et al.~\cite{Ozsoy14}. Matrix factorization methods use the low-rank approximation of input data \cite{Ma:2011:RSS:1935826.1935877}. Example matrix factorization-based recommendation methods belong to Pan et al.~\cite{pan2008one}, Hu et al.~\cite{hu2008collaborative}, Rendle et al.~\cite{rendle2009bpr}, Gao et al.~\cite{gao2013exploring}, Zhang et al.~\cite{zhang2014lore}, Li et al.~\cite{li2015rank}, Zhao et al\cite{ZhaoZYLK16} and He et al.~\cite{HeLLSC16}. Recently deep learning-based techniques have gained more attention from the recommender system domain. The example works utilizing deep learning for recommendation belong to Salakhutdinov et al.~\cite{SalakhutdinovMH07}, Georgiev and Nakov\cite{GeorgievN13}, Wang et al.~\cite{WangWY14}, Musto et al.~\cite{musto2018deep} and Ai et al.~\cite{ai2018learning}. 


Many of the deep learning-based techniques utilize vector space embeddings, e.g., Word2Vec\cite{MikolovSCCD13}, Doc2Vec\cite{LeM14}. Initial methods utilizing vector space embedding methods for recommendation usually use text-based features, like tags or comments, e.g.,~\cite{ShinCL14, MustoSGL15, manotumruksa2016modelling}. More recent ones use non-textual data, such as historical data of purchases or visits to locations. For example, in order to recommend the next purchase item \cite{grbovic2015commerce} employed Word2Vec, to recommend venues \cite{ozsoy2016word} employed Doc2Vec and \cite{liu2016exploring} employed Word2Vec and C-WARP loss (SG-CWARP method), to recommend location \cite{zhao2016gt, zhao2017geo} used Word2Vec (SEER method). There are also methods in the literature which incorporate contextual information while learning the vector space embeddings of the items. To make location recommendation, \cite{zhao2016gt} incorporated temporal (T-SEER) and geographical information (GT-SEER) to their base SEER algorithm and \cite{yang2018unsupervised} incorporated geographical, temporal and categorical information to their Word2Vec-based STES method. To make product recommendations, \cite{guo2018exploiting} utilized sequentiality of the items and used a network embedding technique and collaborative filtering. 


All the aforementioned methods use more traditional embedding techniques, e.g., Word2Vec or Doc2Vec, which cannot learn the representation of the inputs which are not encountered during the training. Also, only a few of the deep learning-based recommendation methods utilize the sequentiality of check-ins \cite{zhao2016gt, guo2018exploiting}. In this paper, I group the items by using the sequentiality of the interactions and extract the semantic relations among the items by FastText, which is powerful at learning the representations. 

 

 





\section{Recommending Sequences}\label{appDeepRec}



In this section, the FastText method and how it is used for recommending a list of sequences is explained.


\begin{figure}
 \centering
 \includegraphics[width=\linewidth]{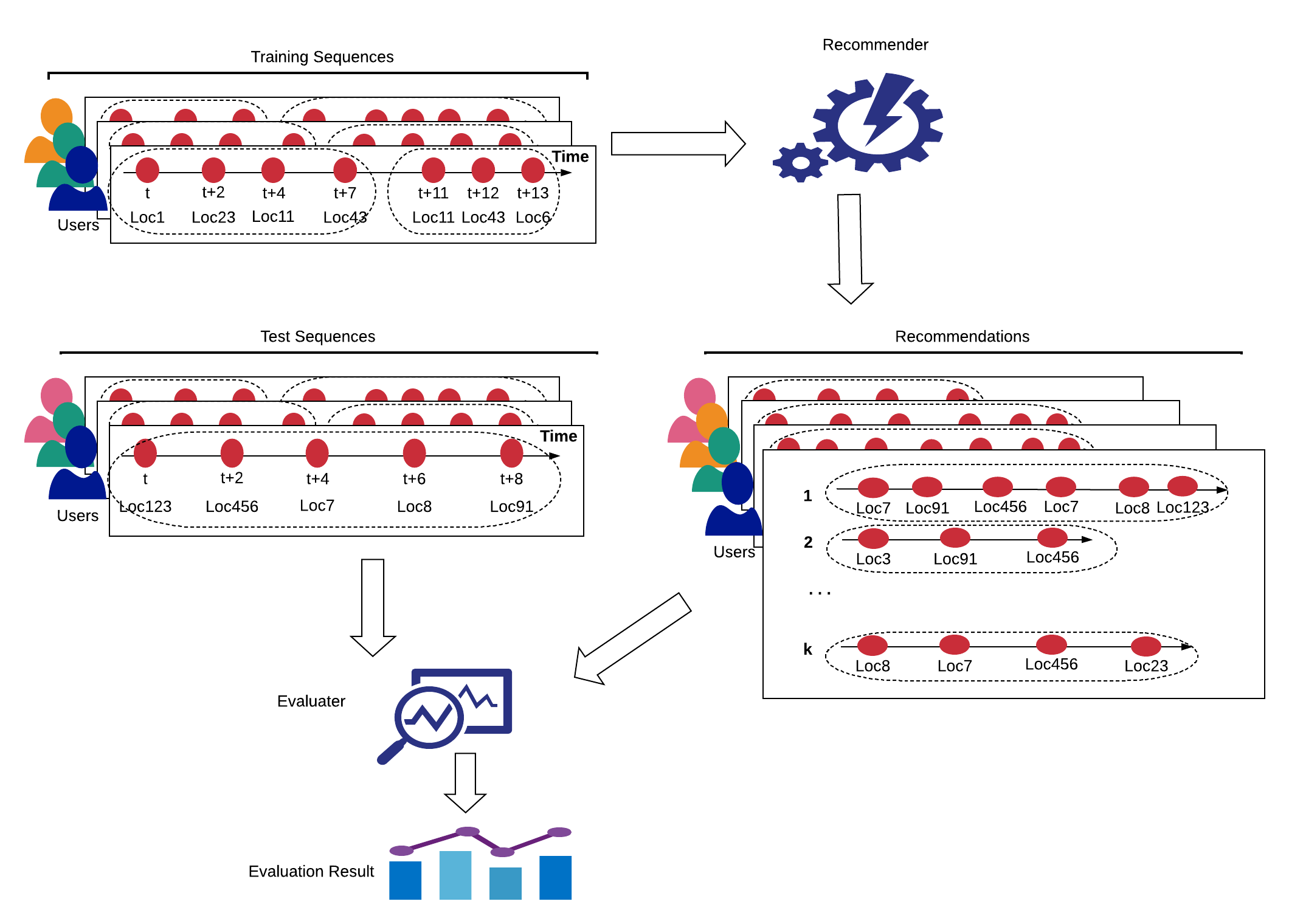} 
 \caption{The high-level sequence recommendation process}
 \Description{overall}
 \label{overall}
\end{figure}


\subsection{The FastText method}\label{fastText}

Vector space embedding models represent the documents and words as vectors to capture the contextual and semantic relations among these textual units \cite{MikolovCoRR13}. Word2Vec \cite{MikolovCoRR13, MikolovSCCD13}, a commonly used vector embedding method in the literature, uses words as the input textual units. FastText method \cite{bojanowski2016enriching} extends the Word2Vec by utilizing the character n-grams of words.

Given the input sentences (sequence of words), Word2Vec captures the semantic and syntactic information of the words and produces low dimensional continuous space representations of them \cite{MikolovCoRR13, MikolovSCCD13, LiXTJZC15}. Word2Vec uses the words as they appear on the input; i.e., without any morphological analysis; learns the representation of the words existing in the training data only and produces different vectors for words even if they share common roots. As a result, in the execution time, Word2Vec cannot return any representation for an unseen word. To overcome these limitations, Bojanowski et al.~\cite{bojanowski2016enriching} proposed the FastText method, which extends the Word2Vec and takes the subword units (character n-grams) into account. 

\begin{figure}
 \centering
 \includegraphics[width=\linewidth]{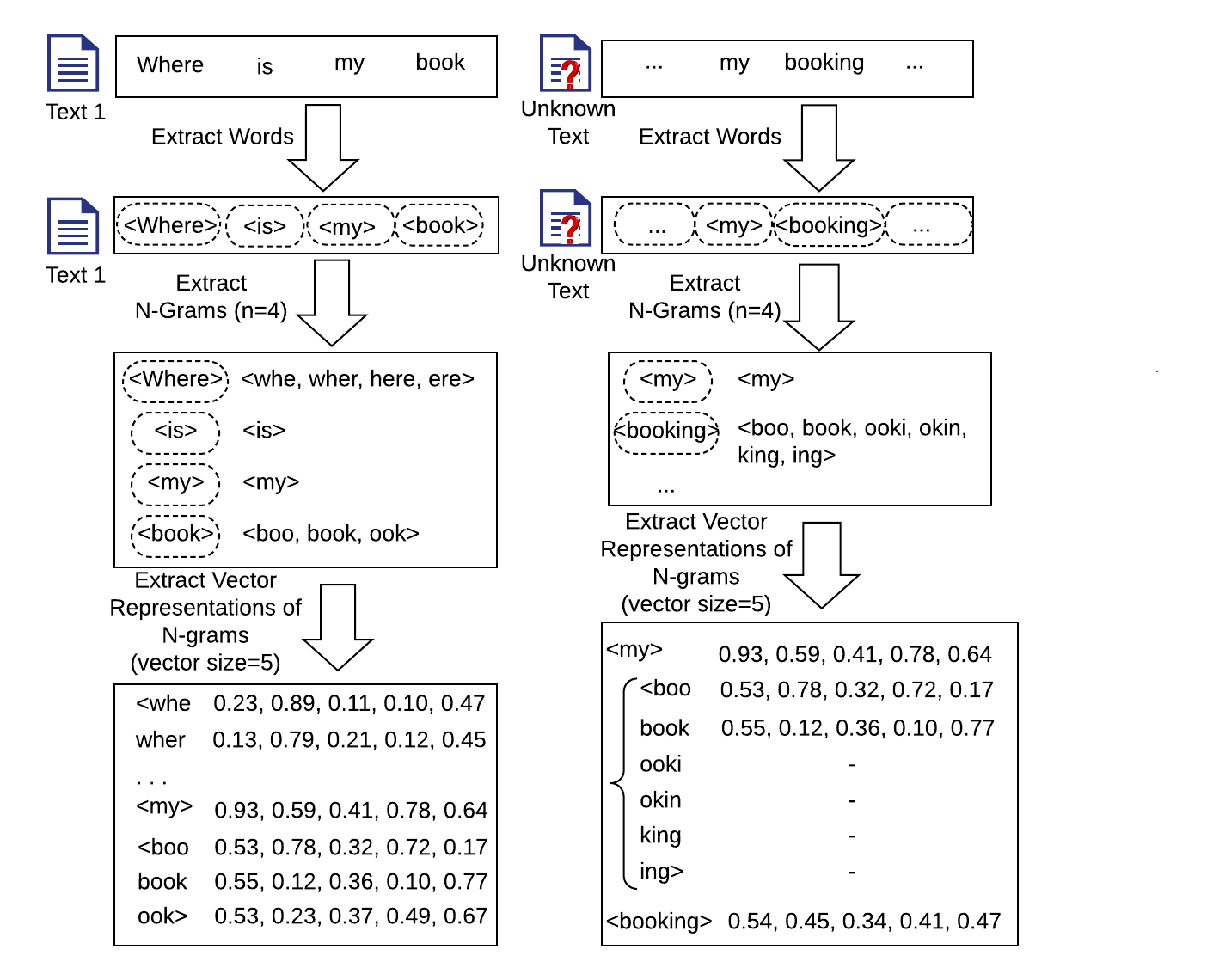} 
 \caption{Example showing how FastText method works}
 \Description{Example showing how FastText works}
 \label{fastText_example}
\end{figure}
The steps of FastText method in training and execution time are presented in the left and right side of the \autoref{fastText_example}, respectively. In the training time, the FastText method models each input word as a bag of character n-grams and produces the vector representations of these n-grams as well as the input words. The vector representations of the input words are calculated by combining the learned vectors of the word itself and its n-grams. In the execution time, the queried word is either (i) already seen in the training data, e.g., the word 'my' in \autoref{fastText_example} or (ii) unseen or a new word, e.g., the word 'booking' in \autoref{fastText_example}. In the former case, the output vector representation of the word is directly returned. In the latter case, the queried word's vector representation is approximated by taking the average of the vectors of its character n-grams.

\subsection{Recommending sequences using the FastText method} \label{fastTextCheckin}
The proposed recommendation method has two main steps: 1) Learning the vector space embeddings of (sub-sequence of) interactions using the FastText method 2) Using the learned vector representations to recommend a list of sequences. 

\subsubsection{Learning the vector space embeddings of interactions} \label{fastTextCheckin_embedding}
In this work, FastText method from natural language processing (NLP) is adapted to the recommender system (RS) by making an analogy in between textual data and user-item interactions data. \autoref{analogy} summarizes the analogy, where textual data (e.g., sentence, word) is mapped to interaction data (e.g., playlists, tracks). 
\begin{table}
 \caption{The proposed analogy in between textual data and interaction data}
 \label{analogy}
 \begin{tabular}{p{2.5cm}|p{4.5cm}}
  \toprule
  \textbf{NLP} & \textbf{RS}\\
  \midrule
  sentences			&	 all interactions per user, e.g., all check-ins or all tracks \\
   \midrule
  words 		&	 sequence of interactions, e.g., playlists\\
   \midrule
  sub-words (n-grams) 		&	sub-sequences of interactions \\
   \midrule
  character 	& individual interactions, e.g., a single venue or a single track\\
 \bottomrule
\end{tabular}
\end{table}

\begin{figure}
 \centering
 \includegraphics[width=\linewidth]{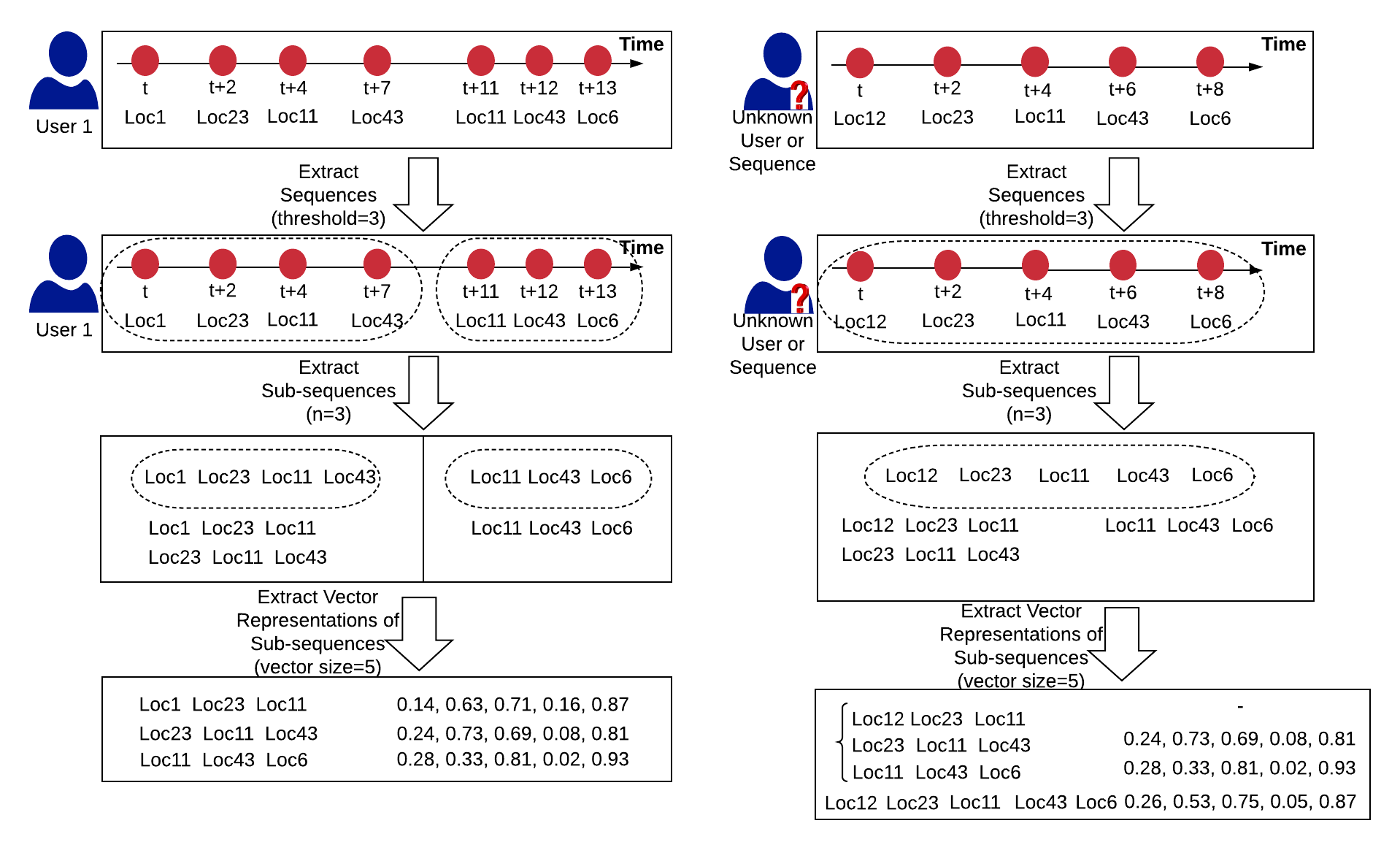} 
 \caption{Example showing how FastText works for recommending venues (locations)}
 \Description{Example showing how FastText works for recommending venues (locations)}
 \label{fastText_rec_example}
\end{figure}

FastText method splits the sequences of input (sentence or interactions) into sequences (words or sequence of interactions) and then forms the sub-units by character n-grams or sub-sequences, as shown in \autoref{fastText_example} and \autoref{fastText_rec_example}, for textual data and interaction data, respectively. Even though check-in data is used in \autoref{fastText_rec_example} as the example, it is possible to use any other interaction data; such as a sequence of music tracks (i.e., playlists) listened or videos watched by the users.

The process of learning the vector space embeddings using FastText method is presented in \autoref{fastTextCheckin_embedding_algo}. Given the all interactions observed in the history of a user, i.e. interaction profile, firstly the sequences are extracted. The sequences can either be self-defined, e.g., as session data or as playlists, or they can be calculated using the temporal features of the dataset. For example,~\cite{guo2018exploiting} groups the items rated in a short time interval together proposing that items rated in a short interval are more likely to be correlated.
The second step of the algorithm is to extract the sub-sequences by grouping n-items together while keeping their temporal order. An example of this process is presented in \autoref{fastText_rec_example}. Then, all the extracted sub-sequences from all of the users are used for training the FastText vector space model. The result of the training process is the vector representations of the input sequences and the sub-sequences. In \autoref{fastText_rec_example}, example of learned vectors of the example check-ins data is presented.


\begin{algorithm}
\SetAlFnt{\small}
\SetAlCapFnt{\small}
\SetAlCapNameFnt{\small}
\KwIn{List of all interactions per user ($L_{All}$)}
\KwOut{Vector space embeddings of sub-sequences of interactions}
$subseq\_L_{All}=\{\}$ \\ 
\ForEach{Sequence $L$ in $L_{All}$}{%
 	 Extract the sub-sequences $subseq_L$ from the $L$\\
	 Collect the extracted $subseq_L$ in $subseq\_L_{All}$} 
Learn the FastText vector space embeddings ($V$) using $subseq\_L_{All}$\\
\Return Vector space embeddings of the seqences and sub-sequences of interactions ($V$)
\caption{{\bf Learning the FastText vector space embeddings of interactions} \label{fastTextCheckin_embedding_algo}}
\end{algorithm}





\subsubsection{Recommending list of sequences}\label{fastTextCheckin_recommendation}
The second step of our proposed method is to recommend a list where each element is a sequence. Even though it is possible to use any kind of algorithm in this step, I use content-based filtering method which is a simple, yet efficient method. In the future, I will combine the proposed method with the state-of-the-art, such as matrix factorization-based or neural networks-based, algorithms.

The traditional content filtering method uses predefined features describing the items. Instead of that, in our proposed algorithm, I utilize the pre-calculated FastText vector representations learned in the previous step. I use the similarities among the items (sequences) to make recommendations, as in the content-based filtering. The process of recommending a list of sequences is presented in the Algorithm \autoref{fastTextCheckin_recommendation_algo}. 



\begin{algorithm}
\SetAlFnt{\small}
\SetAlCapFnt{\small}
\SetAlCapNameFnt{\small}
\KwIn{FastText vector space embeddings ($V$) extracted in the previous step, List of observed sequences of interactions per user ($L_{All}$)}
\KwOut{List of sequences as the recommendation per user}
$user\_2\_recommendations=\{\}$ \\ 
\ForEach{Observed sequence of interactions of a single user $L$ in $L_{All}$}{%
 	Compute the vector space embeddings of $L$ using the FastText method\\
	Compute most similar sequences to $L$ together with the similarity scores using $V$\\
	Decide on the top-k most-similar sequences $rec$ \\ 
	Collect $rec$ in $user\_2\_recommendations$} 
\Return $user\_2\_recommendations$
\caption{{\bf Making sequence recommendations using the FastText vector space embeddings} \label{fastTextCheckin_recommendation_algo}}
\end{algorithm}



Given the observed sequences of a user, the algorithm, firstly, extracts the vector space embedding of the input. The embedding is either (i) directly returned, if the related sequence is observed during the training or (ii) calculated from the known vectors of the sub-sequences. For example, in \autoref{fastText_rec_example}, the vector representation of the input sequence \textit{[Loc12, Loc23, Loc11, Loc43, Loc6]} is computed from the vectors of its sub-sequences. Having the vector representation of the input sequence, the most similar sequences (i.e., top-k) are extracted and recommended.
Unlike traditional recommendation methods, e.g. content-based recommendation, the output of this step is the list of most similar sequences, not the list of individual items. For example, for the check-ins and learned vectors shown in \autoref{fastText_rec_example}, lets assume that the most similar three sequences and their similarity scores are: (\textit{[Loc1, Loc23, Loc11, Loc43], 0.99}), (\textit{[Loc11, Loc43, Loc6], 0.97}) and (\textit{[Loc11, Loc43, Loc64], 0.87}). If the output is a recommendation list with two elements (k=2) will be returned, then the output will be \textit{[[Loc1, Loc23, Loc11, Loc43], [Loc11, Loc43, Loc6]]}, i.e., a list with the two most similar sequences.

 
 


 \section{Evaluating Sequences}\label{eval_rouge}

There are several evaluation metrics for measuring the performance of recommendation algorithms on several different aspects~\cite{gunawardana2015evaluating}. Depending on the purpose of the recommendation algorithm, the offline evaluation metrics aim to measure either prediction accuracy, e.g., RMSE, or classification accuracy, e.g., precision, or rank accuracy, e.g., NDCG,~\cite{herlocker2004evaluating}. 
Additional to accuracy-based metrics, it is possible to measure coverage, novelty, diversity or serendipity of the recommendation algorithms. 
Recently, these established metrics are extended and adapted for evaluating a sequence-based recommender system~\cite{monti2019sequeval}. However, these methods, even the extended versions, consider the sequences as a list of items and they discard the order among the recommended items~\cite{monti2019sequeval}.

\begin{figure}
 \centering
 \includegraphics[width=0.9\columnwidth]{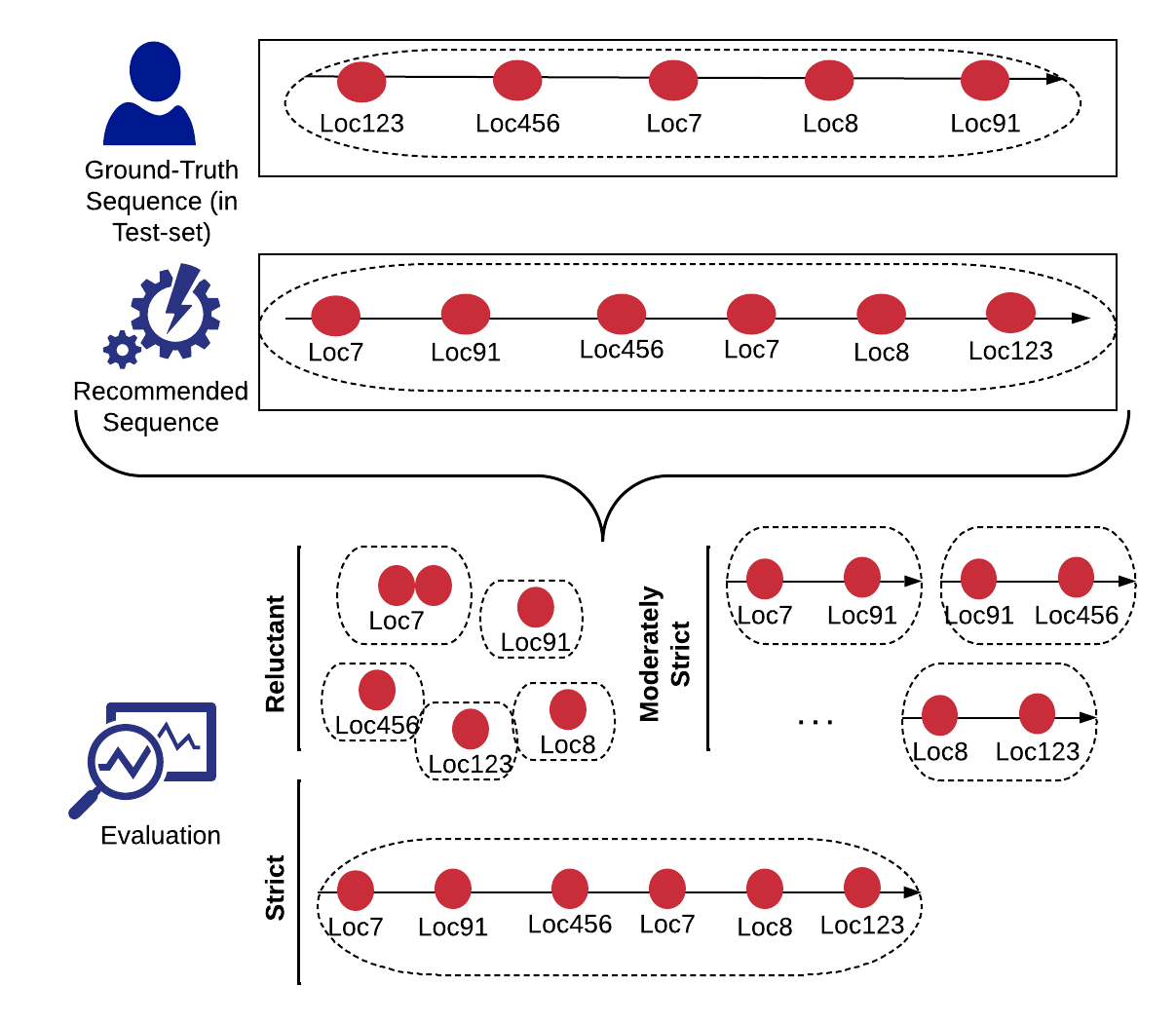} 
 \caption{Example ground-truth and recommended sequence. While reluctant evaluation tends not to take the order in the sequences into account, strict evaluation looks for the exact match. A moderetly strict evaluation metric, like ROUGE score, can balance out each approach.}
 \Description{eval seq.}
 \label{eval_seq}
\end{figure}

It is more complicated to evaluate the performance when sequences are recommended rather than individual items. 
In order to measure the performance of a sequence-based recommender algorithm, a metric which is capable of comparing sequences rather than individual elements is necessary. Let's first observe why this is necessary on an example: In our example, shown in \autoref{eval_seq}, during the test period the user makes a trip; i.e., visits a sequence of locations, and the recommendation method recommends a trip. For the evaluation, I want to find out how much the ground truth trip from the test period and the recommended trip match. 
If I prefer a reluctant evaluation process where I discard the formation of the sequences and use each item in the sequence as the recommendation, then all the target and the predicted items would be same (i.e., \textit{Loc123, Loc456, Loc7, Loc8} and \textit{Loc91}) and the evaluation result would be the highest score. If a very-strict evaluation process where the sequence is a non-decomposable individual element is preferred, there wouldn't be any match on the example recommendation. In this case, the evaluation score would be $0.0$, even if there are some matching sub-sequences, e.g., the bigram \textit{[Loc456, Loc7]} is a matching sub-sequence.



In order to evaluate a sequence-based recommender algorithm, I prefer moderately strict evaluation process, where I should still work on each sequence and I should be able to capture the information provided by the sub-sequences. Luckily, this evaluation problem is also seen in another domain, namely natural language processing (NLP). Creating a sequence of natural language words and sequence of recommended items are strongly similar and metrics from natural language processing (NLP) can be used for the evaluation of sequence-based recommender algorithms~ \cite{monti2019sequeval}. 

Recall Oriented Understudy for Gisting Evaluation (ROUGE) score~\cite{lin2003automatic,lin2004rouge}, which is originally used for evaluating text summarization methods, compares the overlapping n-grams, word sequences, and word pairs between the automatically generated summaries and the ideal summaries created by humans~\cite{lin2004rouge}. I use ROUGE to compare the generated recommendations with the observed sequences provided in the test set. 
There are four different ROUGE measures~\cite{lin2003automatic,lin2004rouge}: ROUGE-N uses n-gram overlaps, ROUGE-L uses the longest common subsequences, ROUGE-W is the weighted version of ROUGE-L and ROUGE-S uses skip-bigram co-occurrence statistics. In this work, I use ROUGE-N and ROUGE-L scores. For the ROUGE score calculations for the sequence-based recommendation, I continue to use our analogy presented in \autoref{analogy}. 

The precision and recall for ROUGE-N and ROUGE-L are calculated by \autoref{eq:rouge_n_recall} - \autoref{eq:rouge_l_precision}, respectively. In the equations r is one of the reference sequence in the set of reference sequences R, s is the recommended sequence, $\text{gram}_{r}$ $\text{gram}_{s}$ are the n-grams (sub-sequences), the \textit{$\text{count}$} indicates the count of the n-grams, \textit{$\text{match}$} indicates the count of the overlapping n-grams, $|r|$ and $|s|$ are the length of the input sequences and \textit{$LCS$} indicates the length of the longest common sub-sequence. LCS does not require consecutive matches but in-sequence matches\cite{lin2004rouge}. F-measure of both ROUGE-N and ROUGE-L scores can be calculated from the recall and precision scores using the traditional method in information retrieval. For the example given in \autoref{eval_seq}, the ROUGE-2 scores will be $precision=0.40, recall=0.50, fmeasure=0.44$ whereas ROUGE-L scores will be $precision=0.50, recall=0.60, fmeasure=0.54$.

\begin{equation}
\label{eq:rouge_n_recall}
\text{ROUGE-N} _{recall} = 
\frac
{\sum\limits_{r \in R}
\sum\limits_{\substack{\text{gram}_{r} \in r, \\ \text{gram}_{s} \in s}} \text{match}(\text{gram}_{r}, \text{gram}_{s})}
{\sum_{r \in R}\sum\limits_{\text{gram}_{r} \in r}\text{count}(\text{gram}_{r})}
\end{equation}

\begin{equation}
\label{eq:rouge_n_precision}
\text{ROUGE-N} _{prec.} = 
\frac
{\sum\limits_{r \in R}
\sum\limits_{\substack{\text{gram}_{r} \in r, \\ \text{gram}_{s} \in s}} \text{match}(\text{gram}_{r}, \text{gram}_{s})}
{\sum\limits_{\text{gram}_{s} \in s}\text{count}(\text{gram}_{s})}
\end{equation}

\begin{equation}
\label{eq:rouge_l_recall}
\text{ROUGE-L} _{recall} = 
\frac
{\sum\limits_{r \in R} LCS(r, s)}
{\sum\limits_{r \in R} |r|}
\end{equation}

\begin{equation}
\label{eq:rouge_l_precision}
\text{ROUGE-L} _{prec.} = 
\frac
{\sum\limits_{r \in R} LCS(r, s)}
{|s|}
\end{equation}


Besides ROUGE metric, there are other evaluation metrics from non-RS domains that can be used for the evaluation of sequence-based recommendation systems. For example, BLEU score~\cite{papineni2002bleu} is used for evaluating machine-translation systems and measures the correspondence between the output translation and the baseline human translation. In this paper, application and comparison of these kinds of non-RS metrics to sequence recommendation is considered as future work.


\section{Evaluation}\label{eval}

This section presents the evaluation configurations; e.g., dataset, metrics, baselines; and the evaluation results.

\subsection{Evaluation Configurations}\label{dataset_evalMetrics_expParameters}

\subsubsection{Datasets}
The proposed method can be used in different domains, e.g., music, movies, check-ins and with different sizes of datasets. Since sequence information is required, datasets which already contain session-like information, e.g., session for online retail websites, playlists for music applications, are preferred. For the experiments, \textit{LearNext} dataset, which contains location-based sequences, and \textit{30Music} dataset, which contains track playing sequences, are used.
 
LearNext dataset~\cite{baraglia2013learnext} contains tourists' movements which are collected from the photos shared on Flickr. The collected data are from three different cities and have different sizes: small (Pisa), medium (Florence), and large (Rome). The movements of each user, i.e., their visits to various locations, are converted into trials by executing the time-based cutting method. The trials provide the user-location relations and ordered list of locations that each user visits. In our experiments, the large (Rome) dataset is used.


30Music dataset~\cite{turrin201530music} is a collection of listening and playlists data retrieved from Last.fm. It is composed of entities; such as users, tracks, artists; and relations; such as play events, play sessions.
The dataset organizes the play events are as listening sessions. When a user listens to a track, the user-track relationships and the play-timestamp order of tracks are saved. This lets us use the sessions information as the input sequence to our system. 




\subsubsection{Training and testing procedure}
\begin{figure}
 \centering
 \includegraphics[width=0.9\linewidth]{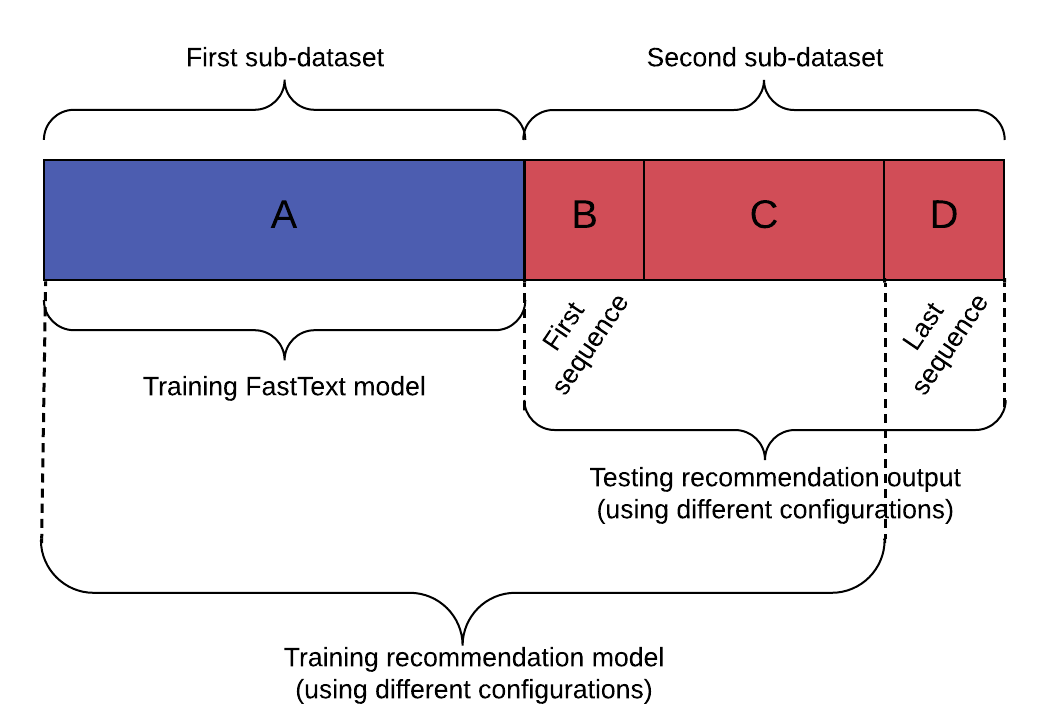} 
 \caption{The configurations for training and testing the methods}
 \Description{The configurations for training and testing the methods}
 \label{eval_piece}
\end{figure}

The dataset is initially divided into two sub-sets, one for training FastText model and the other for training recommendation model and testing. The first sub-set is shown with A in \autoref{eval_piece}. The second sub-set is further divided into three, shown as B, C, D in \autoref{eval_piece}. The sub-set B contains only the first sequence per user (e.g., the first playlist a user has listened), the subset D contains only the last sequence per user and the subset C contains the remaining sequences.

\begin{table}
 \caption{The training and testing configurations of the recommendation models and their outputs}
 \label{eval_config}
 \begin{tabular}{c|c|c}
  \toprule
   & \textbf{Training} & \textbf{Testing}\\
  \midrule
  I	&	 A & B, C, D \\
   \midrule
  II	&	 A, B & C, D\\
   \midrule
  III	&	 A, B, C & D \\
   \midrule
  IV	&	 B & C, D\\
     \midrule
  V	&	 B, C & D \\
 \bottomrule
\end{tabular}
\end{table}

In order to train the FastText model, I always used the sub-dataset A. Whenever further training is required, e.g., for finding neighbors for a KNN-based collaborative filtering, I experimented with different configurations. These configurations are listed in \autoref{eval_config}. For example, for the Evaluation Configuration-II, the subsets A and B are used for the training, and C and D are used for the evaluation. The configurations shown in \autoref{eval_config} let us analyse the efficieny of the recommendations from different perspectives, e.g., for cold-start users.

\begin{table*}
 \caption{FastText-Seq evaluation results (LearNext-Rome)}
 \label{fastTextResults_lnrome}
 \begin{tabular} {c|c|c|c|c}
  \toprule
  						 & ROUGE-1 & ROUGE-1 &  ROUGE-L &  ROUGE-L \\
  Eval. Configuration & precision & recall & precision & recall\\
  \midrule
	 Type I & 0.0122	 & 0.0181 &	0.0122	 & 0.0179 \\
	 Type II & 0.0795 & 0.0904	& 0.0774 &	0.0887 \\
	Type III & 0.0772 & 0.0978	& 0.0750 & 0.0960 \\
	Type IV & 0.0806	& 0.0918 & 0.0785 & 0.0904 \\
	Type V & 0.0767	& 0.0979	& 0.0744& 0.0962\\
 \bottomrule
\end{tabular}
\end{table*}

\begin{table*}
 \caption{FastText-Seq evaluation results (30Music)}
 \label{fastTextResults_30music}
 \begin{tabular} {c|c|c|c|c}
  \toprule
  						 & ROUGE-1 & ROUGE-1 &  ROUGE-L & ROUGE-L\\
  Eval. Configuration & precision & recall & pecision & recall\\
  \midrule
	 Type I & 0.00510	& 0.00982	& 0.00436	& 0.00843	 \\
	 Type II & 0.00714 	& 0.01179	& 0.00611	& 0.01010	\\

 \bottomrule
\end{tabular}
\end{table*}

\subsubsection{Evaluation Metrics}

The evaluation metrics for a sequence-based recommender should be neither too reluctant 
nor too strict.
The unexpected outcomes of both of these approaches are explained and exemplified in \autoref{eval_rouge}. As explained in \autoref{eval_rouge}, ROUGE is used as the main evaluation metric.

\subsubsection{Parameters}

In order to learn the vector representations of venues, the FastText implementation in $gensim$ toolbox is used. While learning the vector representations, default parameters used, except the following three parameters: the length of n-grams, $max\_n$, and the vector size, $s$. 
\begin{itemize}
\item $sg$: Type of the training algorithm (Skip-gram or CBOW). I used both algorithms for the experiments. 
\item $max\_n$: Maximum length of character n-grams. I used values in the range $[1,10]$ with increment of 1 (Keeping $size=100$).
\item $size (s)$: Size of the word vectors. I experimented using the following values: $[10, 50, 100, 150, 200, 250]$ (Keeping $max\_n=5$).
\end{itemize}
To make recommendations by the learned vector representations, I used the number of neighbors ($N$) and output list size ($k$) parameters and both are set to 10 for the experiments. 





\subsubsection{Baselines}

\begin{figure}
 \centering
 \includegraphics[width=\columnwidth]{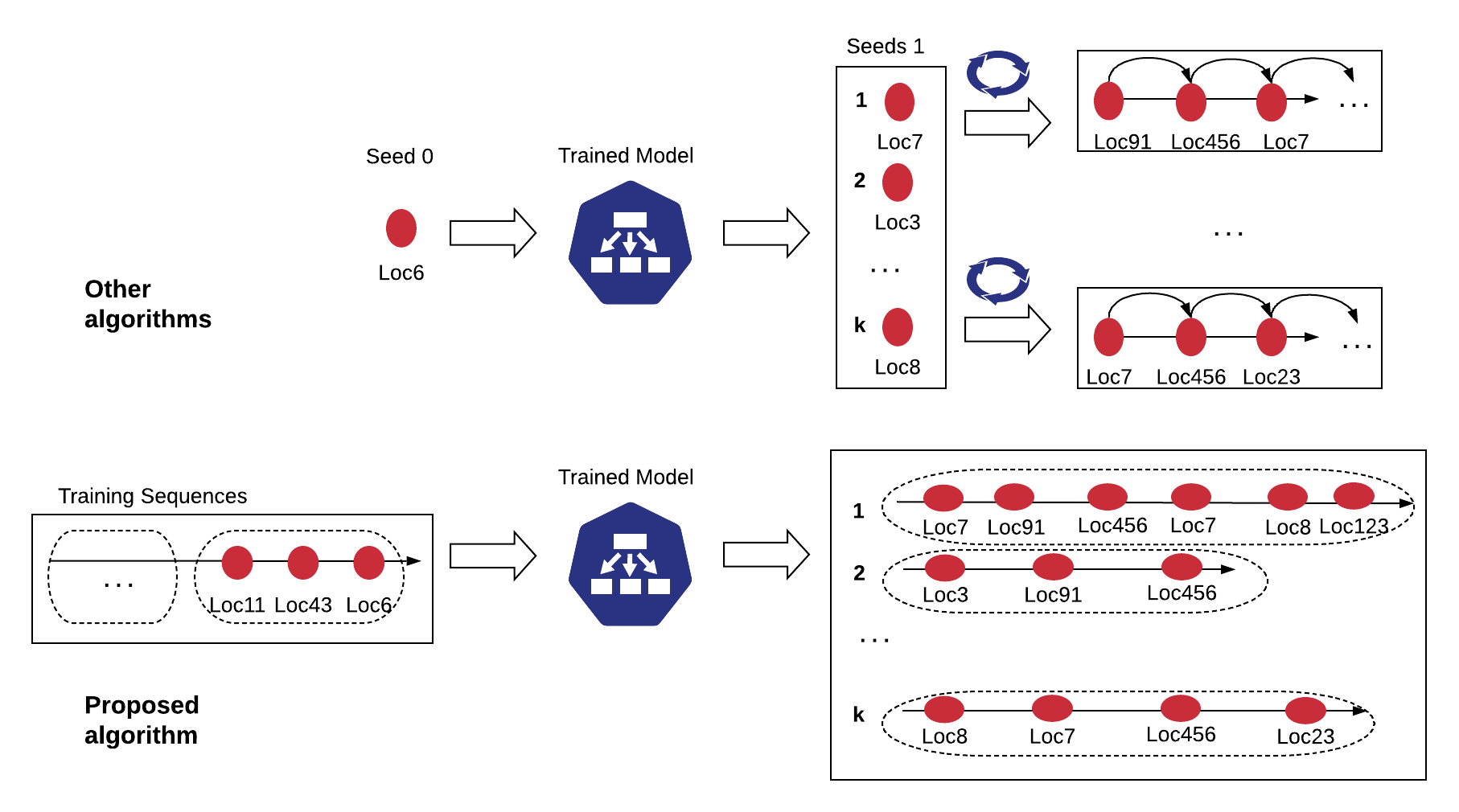} 
 \caption{Recommending list ofmulti-item sequences (e.g., list of playlists)}
 \Description{figure for multi seq. production}
 \label{multi_item_seq}
\end{figure}

In RS literature, there are various algorithms which may/not take the sequence information into account. Even the sequence-based methods usually focus on recommending next-item only, and they are not capable of recommending longer sequences. On the contrary, the proposed method is capable of recommending longer sequences (\autoref{multi_item_seq}). 


\subsection{Evaluation Results}\label{evalResults}

\


In this paper, the effect of utilizing the FastText method is evaluated by using ROUGE metrics. For both of the LearnNext-Rome and 30Music datasets., the best performances are obtained when $s=128$ and $max\_n=9$. 
\autoref{fastTextResults_lnrome} and \autoref{fastTextResults_30music} shows the evaluation results for the LearnNext-Rome and 30Music datasets, respectively.

The evaluation results reveal that Evaluation Configuration Type-I performs worse than the others. In this evaluation configuration, only the first half of the dataset is used during the training (of the FastText model and recommendation model). 
The first half of the datasets (shown in \autoref{eval_config}) contain information from some of the users, such that there are many users who start to use the system in the second half. These users can be considered as cold-start users. Not having any information from those users reduces the performance of the recommender system. However, when the cold-start user interacts with the system once (as in Evaluation Configuration Type-II), the performance of the recommender increases(by nearly 50\% for 30Music and even more for LearNext-Rome datasets.).

Further experiments on the LearNext-Rome dataset reveals that while making a recommendation not using the data in the first half, which is used for training the FastText, does not affect the results much. i.e., Evaluation Configuration Type-IV/V. Also, evaluation results reveal that the proposed algorithm is efficient not only at recommending a list of sequences (Type I/II/IV) but also a single sequence (Type III/V).

\section{Conclusion}\label{conclusion}
Vector space embeddings and deep learning methods are commonly used in Recommender Systems (RS) domain~\cite{grbovic2015commerce, musto2018deep, ai2018learning,yang2018unsupervised}. However, most of these recommender methods overlook the sequentiality feature and consider each interaction, e.g., check-in, independent from each other~\cite{zhao2016gt, guo2018exploiting}. 

In this work, a method is proposed to consider the sequentiality of the interaction of users with items. The proposed method uses a technique from the natural language processing (NLP) literature where sequentiality naturally occurs, e.g., word order. The proposed method uses FastText~\cite{bojanowski2016enriching} to model the relationship among the units of the sequences, e.g., tracks, playlists, provides that information as an input to a traditional recommender method and recommends a list of multi-item sequences. In addition to modelling and recommending a list of multi-item sequences, this paper proposes to use an evaluation metric from NLP, namely ROUGE~\cite{lin2003automatic,lin2004rouge}, to evaluate the effectiveness of the recommended sequences.
The current experimental results reveal that it is possible to recommend a list of multi-item sequences, in addition to the traditional next item recommendation. The usage of FastText, which utilise sub-units of the input sequences, helps to overcome cold-start user problem.

Even though current experimental results are promising, there are many missing pieces in the experimental section. In the future, I want to analyse and execute experiments on (i) the performance when other metrics are used, e.g., precision, novelty, (ii) the effects of the hyperparameters, e.g. embedding length and (iii) the comparison to the baseline algorithms. Also, in the future I want to explore more recent, transformer-based algorithms on the same problem, namely recommending a list of multi-item sequences.

%
\bibliographystyle{ACM-Reference-Format}
\bibliography{MyTex}

%

\end{document}